# SAD effects on grantsmanship

*George A. Lozano*

Estonian Centre of Evolutionary Ecology

Tahe 15

Tartu, Estonia, 50108

email: dr.george.lozano@gmail.com

**Seasonal affective disorder (SAD) is a state of depression induced by a lack of sufficient sunlight that occurs at high latitudes during the fall and winter. One effect of SAD is that causes people to be more risk-adverse, an effect that should be considered by granting agencies of high latitude countries. Funding agencies often have programmes aimed at high-risk, innovative research. However, the time of the year during which these purposefully high-risk proposals are evaluated usually does not take into consideration the effects of SAD. In high-latitude countries (e.g., Canada, UK, Nordic and Baltic countries), evaluating proposals for high-risk programmes during the late fall might significantly detract from the very purpose of such programmes. At this time of the year, grant evaluators might be in a darkness-induced state of mild depression. As such, evaluators might be more likely to opt for safe investments, more of the same, the well established, which is the antithesis of innovative research.**

Seasonal affective disorder (SAD) is a state of depression induced by a lack of sufficient sunlight that occurs in people living at high latitudes during the fall and winter. The symptoms are typical of most forms of depression. Physical symptoms include lethargy, inability to concentrate, increased sleeping and eating, lack of activity, and weight gain. Psychological symptoms include irritability, sadness, pessimism, guilt, worthlessness, lack of creativity, low sex drive, increased thoughts of suicide, and loss of interest in hobbies and social activities [1,2]. SAD can affect up to 10-15% of the population, and it is generally more prevalent at higher latitudes [3]. Sub-clinical symptoms occur in at least as many people [4], so clinical SAD is just a extreme case of an otherwise relatively common condition.

Depression leads to risk-aversion in investment decisions [5]; SAD-induced depression has the same effect. For instance, seasonal cycles in stock markets, and a 6 month shift between market cycles in the 2 hemispheres, are both consistent with the hypothesis that investors are most risk-averse in the fall and winter, and most risk-prone in the spring and summer [6]. Similarly, initial public offerings (IPOs) in the U.S.A. are relatively underpriced in the fall and winter compared to in the summer and spring [7]. Also, stock analysts in the USA are significantly less optimistic about their forecasts during SAD months than the rest of the year, more so in northern states [8]. In a laboratory setting, people who are affected by SAD are significantly more likely than control individuals to avoid risk in financial decisions, but only during the winter [9]. Fluctuations in the stock market are driven by individual decisions of million of people; hence, it is apparent that clinical and sub-clinical forms of SAD are strong enough to affect decision-making in the general population.

One might think that this information is important only to investors and CEOs, but at least one other group should heed these results: granting agencies in high-latitude countries. Investing and grant evaluation are similar processes. In both cases, assessors, whether they are called investors or evaluators, must choose among many possible ventures. To arrive at a decision, assessors use past performance, current leadership, infrastructure, location, feasibility, collaborators, and partners, and, of course, the expected return on investment money. Unlike investors, evaluators do not directly benefit from their investments; after all, it is not their money, so they are more like investment consultants. As such, they have a fiduciary duty to try to obtain the greatest return (impact) for the grant agency's money [10].

Funding agencies generally are aware that, as in investing, greater risk increases both the potential reward and the possibility of failure. Innovation is essential to scientific progress, but it is also inherently risky. Hence, agencies often have specific programmes aimed at high-risk, innovative research. High-risk funding programmes might specifically cater to risky research projects, people who wish to work outside the comfort of their usual research groups or fields, individuals as opposed to groups, and tangential ideas and projects. In contrast, low-risk funding programmes might include continuing funding, institutional funding that must be awarded, and funding to large or politically well-connected groups that, realistically, would not be denied funding.

National granting agencies invite, accept, assess, and fund research proposals throughout the year. By necessity, some programmes are tied to specific times of the year, sometimes linked to academic or fiscal calendars, and other programmes are scattered across the year. However, the timing of purposefully high-risk programmes usually does takes into account the fact that proposal writers, and perhaps more importantly, the expert reviewers who evaluate the proposals, and agency officials who eventually make the funding decisions, might all be influenced by SAD.

Granting agencies in high latitude countries must be careful to time these "high-risk", "innovative research" competitions such that they are not written or evaluated during the SAD months. During the late fall, November and December, as days become increasingly shorter, people might very well be walking around in a darkness-induced state of mild depression. As such, they would be more likely to opt for safe investments, more of the same, the well established, which is the antithesis of innovative research. In high-latitude countries, submitting and, perhaps more importantly, evaluating high-risk, innovative, proposals during the late fall might significantly detract from the very purpose of such programmes.

SAD can affect both the process of reviewing and the act of writing a research proposal. As long as the call for proposals is made sufficiently in advance, there is no way of knowing when exactly the proposals are written and the main ideas are generated. However, a proposal's final review would probably be close to the submission deadline. If that submission deadline is during the SAD months, there is a higher likelihood that the proposal writers would remove the bolder, riskier, and perhaps most innovative parts of their proposals. A particularly interesting situation might occur when purposely high-risk proposals are written during the SAD months and evaluated in the spring, or vice-versa. In any case, granting agencies in high latitude countries might have to ensure that neither the evaluation nor the submission deadlines of "high" risk proposals coincides with the SAD months.

In confirming this hypothesis, one must assess both the mean impact of the funded proposals, and the variance of that impact. Proposals funded by high risk programmes would, by design, include atrociously dismal failures and surprisingly wonderful successes. Hence, if the programme is serving its purpose, the mean impact might be the same, but the variance would be higher. A possible test of this idea might involve comparing instances when the same competition is conducted at two times of the year, and assessing the eventual impact of the funded proposals. Alternatively, one could search for instances when then timing of the evaluation and/or submission deadline of a particular programme has changed, and examine the impact of the said research before and after the change. In the lab, it is possible to carry out controlled experiments whereby proposals could be evaluated at different times of the year by the same, or by different people. As always, each approach has distinct strengths and weaknesses.

The effects of SAD have been documented in investing, but not yet in grantmanship. However, just like entrepreneurs and investors, researchers and evaluators are not necessarily immune to the effects of SAD. If the effects are similar, and there is no reason why they should not be, it would be easy for granting agencies change the of submission and evaluation times for competitions for research that is specifically meant to be high-risk, just to be safe.

## Acknowledgements

I thank the University of Tartu for providing me free access to their online library. I thank Drs. M. Kery, A. Ros, and G. Paz-y-Miño for the comments on previous versions of this manuscript. This is contribution number 1412 of the ECEE (reg. no. 80355697).